\documentclass[conference,10pt,twocolumn]{IEEEtran} 
\usepackage{setspace}
\usepackage{cite} 
\usepackage{graphics}
\usepackage{graphicx}
\usepackage[english]{babel}
\usepackage{amsmath}
\usepackage{amsthm}
\usepackage[table,xcdraw]{xcolor}
\usepackage{subfig}
\usepackage{subcaption}
\usepackage[linesnumbered, ruled]{algorithm2e}
\usepackage[noend]{algpseudocode}
\usepackage[font=footnotesize]{caption}
\usepackage{epsfig}
\usepackage{xfrac}
\usepackage{nicefrac}
\usepackage{psfrag}
\usepackage{times}
\usepackage{balance} 
\usepackage{amssymb}
\usepackage{amsfonts}
\usepackage{bm}
\usepackage[bottom]{footmisc}
\usepackage{mwe}
\usepackage{color}
\usepackage{tabularx}
\usepackage{booktabs,multirow}
\usepackage{verbatim}
\usepackage{hyperref}
\usepackage{float}
\usepackage{mwe}
\usepackage{booktabs,multirow}
\usepackage{tikz}
\usetikzlibrary{patterns,arrows}
\usepackage{physics}
\usepackage{verbatim}
\usepackage{wrapfig}
\usepackage{soul}
\usepackage{subcaption}
\usepackage{caption}

\usepackage[left=1.62cm,right=1.62cm,top=1.9cm]{geometry}
\DeclareMathOperator*{\argmax}{arg\,max}
\DeclareMathOperator*{\argmin}{arg\,min}

\SetKwRepeat{Do}{do}{while}
\SetKwInOut{Input}{Input}
\SetKwInOut{Output}{Output}
\SetKw{KwBy}{by}

\usepackage{acronym}
\newacro{UE}{User Equipment}
\newacro{AP}{Anchor Point}
\newacro{CNN}{Convolutional Neural Network}
\newacro{FCNN}{Fully Connected Neural Network}
\newacro{GNSS}{Global Navigation Satellite System}
\newacro{DA}{Data Augmentation}
\newacro{ML}{Machine Learning}
\newacro{DL}{Deep Learning}
\newacro{RNN}{Recurrent Neural Network}
\newacro{NN}{Neural Network}
\newacro{CSI}{Channel State Information}
\newacro{GAN}{Generative Adversarial Network}
\newacro{VAE}{Variational Autoencoder}
\newacro{RSSI}{Received Signal Strength Indicator}
\newacro{PDP}{Power Delay Profile}
\newacro{ERM}{Empirical Risk Minimization}
\newacro{RRM}{Regularized Risk Minimization}
\newacro{OFDM}{Orthogonal Frequency Division Multiplexing}
\newacro{MPC}{Multipath Components}
\newacro{LOS}{Line of Sight}
\newacro{NLOS}{Non-Line of Sight}
\newacro{WSSUS}{Wide Sense Stationary Uncorrelated Scattering}
\newacro{MSE}{Mean Square Error}
\newacro{RMSE}{Root Mean Square Error}
\newacro{ACF}{Autocorrelation Function}
\newacro{TL}{Transfer Learning}
\newacro{TOA}{Time-of-Arrival}

\usepackage{titlesec}
\titlespacing*{\section}{0pt}{0.1\baselineskip}{0.2\baselineskip}
\titlespacing*{\subsection}{0pt}{0.1\baselineskip}{0.2\baselineskip}

\newcommand\blfootnotetext[1]{
    \begingroup
    \renewcommand\thefootnote{}\footnotetext{#1}
    \addtocounter{footnote}{0}
    \endgroup
}

\begin{document}

\title{\fontsize{21}{24}\selectfont Wireless Channel Aware Data Augmentation Methods for Deep Learning-Based Indoor Localization}

\author{ 
Omer Gokalp Serbetci, \textit{Graduate Student Member, IEEE}, Daoud Burghal, \textit{Member, IEEE}, \\ Andreas F. Molisch, \textit{Fellow, IEEE}
}
\maketitle

\begin{abstract}
    Indoor localization is a challenging problem that - unlike outdoor localization - lacks a universal and robust solution. \ac{ML}, particularly \ac{DL}, methods have been investigated as a promising approach. Although such methods bring remarkable localization accuracy, they heavily depend on the training data collected from the environment. The data collection is usually a laborious and time-consuming task, but \ac{DA} can be used to alleviate this issue. 
In this paper, different from previously used \ac{DA}, we propose methods that utilize the domain knowledge about wireless propagation channels and devices. The methods exploit the typical hardware component drift in the transceivers and/or the statistical behavior of the channel, in combination with the measured \ac{PDP}. We comprehensively evaluate the proposed methods to demonstrate their effectiveness. This investigation mainly focuses on the impact of factors such as the number of measurements, augmentation proportion, and the environment of interest impact the effectiveness of the different \ac{DA} methods. We show that in the low-data regime (few actual measurements available), localization accuracy increases up to 50\%, matching non-augmented results in the high-data regime. In addition, the proposed methods may outperform the measurement-only high-data performance by up to 33\% using only 1/4 of the amount of measured data. We also exhibit the effect of different training data distribution and quality on the effectiveness of \ac{DA}. Finally, we demonstrate the power of the proposed methods when employed along with \ac{TL} to address the data scarcity in target and/or source environments.
\end{abstract}

\blfootnotetext{O. G. Serbetci and A. F. Molisch are with the Ming Hsieh Department of Electrical and Computer Engineering, University of Southern California (USC), Los Angeles, CA 90089, USA. (Emails: \{serbetci, molisch\}@usc.edu)}
\blfootnotetext{D. Burghal was with the Ming Hsieh Department of Electrical and Computer Engineering at USC, now is with Samsung Research America, Mountain View, CA 94043, USA. (Email: daoud.burghal@outlook.com).}
\blfootnotetext{ Part of this work is presented at IEEE GLOBECOM
2023 \cite{globecom_paper}. This work was financially supported by NSF grant 2003164.}
\section{Introduction} \label{sec:intro}
    \subsection{Motivation} \label{sec:motivation}
        Indoor localization has many important applications, such as tracking patients in hospitals, localization of first responders during emergency operations, and improving wireless service quality using position information. Despite a large amount of research (see Sec. I.B), there is no single and robust solution to the indoor localization problem - this is in marked contrast to outdoor localization,\footnote{Note that outdoor localization can still be challenging for few environments, such as dense urban environments.} for which \ac{GNSS} has become the default approach \cite{zekavat2019handbook,prof_molisch}. However, \ac{GNSS} may be unavailable in indoor scenarios due to the strong signal attenuation; even when it is available, the impact of strong multipath often reduces the accuracy below the point where it is useful for indoor applications \cite{gps}. 

Earlier approaches to indoor localization problems include trilateration and proximity-based methods (such as \ac{TOA}-based methods \cite{zekavat2019handbook,aditya_toa}), which may, however, provide low accuracy, especially in complex environment structures. For these reasons, data-driven approaches, particularly \ac{ML} algorithms, have gained significant interest \cite{survey_localization}. These algorithms construct the mapping between collected data points (channel characteristics) and the corresponding locations; unlike classical approaches, these models are not tied to the physics of propagation or the environment. Random forest and \ac{NN}-based solutions (\ac{CNN}, \ac{RNN}, transformer) are examples of such methods \cite{cnn, csi_attention, forest, rnn}. Two critical aspects of the ML methods are generalizability, i.e., performance over the unseen data points in the given environment, and transferability, i.e., allowing models to perform well in other environments by gathering only a small amount of data in those environments. 

\ac{ML} localization methods differ not only in the above-discussed mapping functions but also in the mapping elements, namely features and corresponding labels. Most of the earlier ML-based solutions used \ac{RSSI} as features. However, the recent interest has shifted towards \ac{CSI}, i.e., complex transfer function and their multi-antenna equivalents \cite{survey_localization}. This can be attributed to  \textit{i)} Most modern wireless systems, including LTE, 5G NR, and WiFi, are wideband and are capable of acquiring \ac{CSI} as part of the communication protocol anyway \cite{prof_molisch,dahlman20205g}. \textit{ii)} \ac{CSI} provides finer granularity than \ac{RSSI}, which can be valuable for the localization problem.\footnote{The actual features used in the ML methods can be either raw or processed measurement information.}
To elaborate on \textit{ii)}, in supervised settings, where each data point has a label, the labeling options depend on the task, with the most common ones being the 2-D/3-D position or the room/apartment in a building. The selected feature and label pairs affect the formulation of the problem (regression vs. classification) and the \ac{ML} algorithms' corresponding performance. 
The underlying reason is that the information contained in the features and their relation to the labels are different; e.g., two locations with a similar RSSI might have completely different CSI. In this paper, we thus focus on CSI-based methods. 

One challenging aspect of ML-based localization solutions is the fact that the constructed models depend on the label feature spaces and the distribution of the data. The latter is complex and varies due to the nature of wireless environments and measurement devices. Thus, an ML method designed for a certain feature-label space may not work well enough for another, so that the model might need to be retrained vironments. One promising solution to reduce the amount of data needed from the new environment and utilize data from prior (alsis tonown as "source") environments is \ac{TL}. In both cases, data usingfrom the target environment are still required. 
Unfortunately, collecting data from each environment is laborious and time-consuming, a problem that cannot be overlooked for \ac{ML} methods, especially \ac{DL} models that are data-hungry. Furthermore, certain areas in a building may be off-limits for data collection either permanently or at certain times, restricting the amount of data that can be collected. 

To mitigate the burden of data collection, \ac{DA} can be used. \ac{DA} exploits domain knowledge about the features and labels to generate new data using the existing dataset. Such methods apply transformations over the data points' features and help to inject external knowledge about the domain into the learning process. \ac{DA} has demonstrated a significant performance improvement compared to un-augmented approaches by overcoming overfitting during training \cite{aug_survey}. \ac{DA} has been widely used in image classification tasks \cite{aug_survey}, where images are rotated, noise is added, or colors are changed, etc., since such manipulations are natural for images. However, such methods are not a natural fit for wireless channels - it is not even clear how some of these techniques could be applied to \ac{CSI} at all. Thus, this paper employs wireless channel and system knowledge to create \ac{DA} approaches that are a natural fit for \ac{CSI}-based localization approaches and consequently significantly reduce the burden of training data acquisition. Moreover, we show how effective such methods are in different environments, data regimes, and scenarios. 
        
    \subsection{Related Work} \label{sec:related_work}
        \textbf{ML-Based Indoor Localization}. A large number of papers have been published on \ac{ML}-based indoor localization; see \cite{survey_localization} and reference therein; the following just points out some particular examples. These papers introduce various \ac{NN} architectures and ML methods ranging from \ac{RNN} \cite{rnn} to  \ac{CNN} \cite{cnn} to attention mechanism \cite{csi_attention}. In addition to the differences in the method selection, there are different feature spaces across the methods, ranging from \ac{RSSI} \cite{rssi} to phase information \cite{phasefi} to time of flight and angles \cite{wild1} or raw \ac{CSI} \cite{csi_attention, deepfi}. Note that we also consider using raw \ac{CSI}, fed to the neural networks, as the features of the data samples. 

\noindent\textbf{Data Augmentation}. In Computer Vision (CV) applications, using \ac{DA} is very common. The \ac{DA} methods include variations and mixtures of color shifting, pixel rotation, and noise injection \cite{aug_survey}. In natural language processing (NLP) tasks, another set of  \ac{DA} methods are used, which include back translation of translated sentences, random synonym changes, or random swaps within input sentences. These types of augmentation methods are called implicit sampling from random transformations. In \cite{thousand}, the authors claim that such transformations act as regularizers and derive an explicit regularizer for the parameter updates during training.

In indoor localization, in addition to noise injection, e.g., as in \cite{gao2022toward}, there are two main approaches to DA: generative and domain knowledge-based. The former methods include \acp{GAN} \cite{GAN,li2019af,wei2021data} and \acp{VAE} \cite{rizk_augs}. However, such architectures are already data-hungry, i.e., they require an abundance of data for stable performance, defying the purpose of \ac{DA}. Instead, domain knowledge of wireless systems can be used. The most common method is the addition of noise to the raw measurement data \cite{rizk_augs}. Recently, other methods have been introduced. \cite{data_aug2} first processes \ac{CSI} into images and then applies noise injection to simulate the actual communication conditions while generating new data. Ref. \cite{data_aug} introduces a dropout-like mechanism for augmentation. Ref. \cite{self_supervised} (a paper written in parallel to, and independent of, our work, compare our conference paper \cite{globecom_paper}) proposes methods that include sub-carrier flipping, random gain offset, random fading component, and random sign change. Differences to our approach include the random gain offset shifts each subcarrier's amplitude independently (our amplitude shifts are imposed on a per-transceiver basis), the random fading is considered in the frequency domain and applied only to the absolute part of the transformed view of channel frequency response, and transformations are applied to the stacked(real, imaginary and absolute) view of channel frequency response, used in embeddings learned via self-supervised training, while our methods operate directly on the raw CSI.

    \subsection{Main Contributions} \label{sec:contributions}
        This paper introduces different \ac{DA} methods based on the characteristics of wireless propagation channels and devices to enhance indoor localization performance. Different from earlier works, the proposed methods are inspired by physical phenomena in wireless channels. Thus, the proposed \ac{DA} mimics the realistic variations of the wireless signals. Furthermore, going beyond the state of the literature, 
this work investigates how the proposed \ac{DA} methods affect the localization performance in various environment types, dataset sizes, augmentation factors, and the spatial distribution of the dataset over the environment of interest. Our contributions are summarized as follows:
\begin{itemize}
    \item We propose \ac{DA} that generates additional data with random perturbation to the phase and amplification of the signal at \acp{AP}, including possible multi-antenna structures. The methods are explained and motivated with realistic transceiver behaviors.    

    \item We propose methods that create different realizations of the small-scale fading consistent with a measured transfer function. In other words, we create channel realizations that will occur in the close vicinity of the measurement point. 
    We introduce \ac{PDP} based methods to exploit potential sparsity and to preserve the essential statistical characteristics of the measured channel. We propose four different methods: \textit{i)} injecting random phase to each delay bin, \textit{ii)} generating Rayleigh fading from the measured local \ac{PDP}, \textit{iii)} generating complex Gaussian (Rayleigh) fading from a \ac{PDP} averaged over a region, and \textit{iv)} injecting random phase to the highest-power delay bin and generating Rayleigh fading in the other bins. 
    
    \item We extensively study the effect of different dataset regimes, namely \textit{i)} low, \textit{ii)} medium, \textit{iii)} high data regimes, clarifying how \ac{DA} is affecting the training process for all of the introduced DA methods. We use four different real-world (measured) datasets to showcase the performance of the proposed methods. 
    
    \item We investigate how the dataset partitioning impacts the localization performance for both (i) areas that require interpolation (which is used in the literature) and (ii) other areas that require extrapolation in the same environment. We also show that some samples are more valuable than others in terms of generalizability and - based on this concept - learn how to augment a dataset better.
    
    
    \item Finally, we study \ac{TL} approaches to show how the proposed \ac{DA} methods work in realistic cases and show the effect of \ac{DA} both in source and target domains in different data regimes and augmentation ratios. This is important because \ac{TL} significantly reduces the data collection burden for new environments by exploiting a pre-trained model. 
\end{itemize}

    \subsection{Paper Organization} \label{sec:organization}
        The rest of the paper is organized as follows: Sec. \ref{sec:background} provides background information about \ac{DL} and \ac{DA}. Sec. \ref{sec:csi_indoor} introduces the transceiver and channel models followed by the \ac{DL} formulations we use throughout the paper, which include the feature and label spaces and the optimization objectives. The section also establishes the overall \ac{DA} problem formulation. Sec. \ref{sec:transceiver} introduces the proposed algorithms based on transceiver behavior, whereas Sec. \ref{sec:channel_algos} explains the algorithms based on the channel model and \acp{PDP} of the measured samples. Finally, Sec. \ref{sec:num_results} discusses the evaluation methodology (dataset, neural network, and optimization parameters) and demonstrates the performance of the algorithms in different scenarios and data regimes. 
        
\section{Background} \label{sec:background}
    \subsection{Deep Learning} \label{sec:deep_learning} 
        In supervised settings, a \ac{DL} algorithm $\mathcal{A}$ is given a dataset $\mathcal{D}$ and aims to find a mapping from a certain hypothesis $\mathcal{F}$, which contains models $f: \mathcal{X} \rightarrow \mathcal{Y}$. Here, $\mathcal{X}$ and $\mathcal{Y}$ are called feature and label spaces, respectively. We call $\boldsymbol{x} \in \mathcal{X}$ a feature and $\boldsymbol{y} \in \mathcal{Y}$ a label. Then, the dataset $\mathcal{D} = \{(\boldsymbol{x}_i, \boldsymbol{y}_i)\}_{i = 1}^N$ consists of $N$ feature-label tuples. A model in the class $\mathcal{F}$ consists of layers containing neurons connected with weights. Each layer consists of a certain type of weight matrices that apply a linear transformation to the input, and a nonlinear activation function follows it. The selection of connection between neurons and corresponding activation function types depends on the hypothesis class $\mathcal{F}$. 

The main objective of the algorithm $\mathcal{A}$ is finding the model from the hypothesis class such that the true risk over the given loss function and data distribution over feature and label space $P_{\mathcal{X}\mathcal{Y}}$ is minimized, where loss function is $\ell: \mathcal{Y}\times \mathcal{Y} \rightarrow \mathbb{R}$. To be more precise: 
\begin{equation}
    f^* \triangleq \argmin_{f\in \mathcal{F}} R(f)
\end{equation} where $R(f) \triangleq \mathbb{E}_{(\boldsymbol{x}, \boldsymbol{y}) \sim P_{\mathcal{X} \mathcal{Y}}}[\ell(f(\boldsymbol{x}), \boldsymbol{y})]$.

The problem with the previous objective is that we do not have access to the data distribution. Still, we have access to the samples from the distribution $P_{\mathcal{X}\mathcal{Y}}$, i.e., the training dataset $\mathcal{D}_{\mathrm{train}}$. \ac{ERM} and \ac{RRM} as learning algorithms are the reigning paradigms for finding models with low true risk \cite{bendavid}. The output model of the ERM algorithm for a dataset with $N$ samples is given as follows:
\begin{equation} \label{eq:ERM}
    \hat{f} \triangleq \argmin_{f\in \mathcal{F}} \frac{1}{N}\sum_{i=1}^N\ell\left(f(\bold{x}_i), \bold{y}_i\right)
\end{equation}

The model found in Eq. (\ref{eq:ERM}) $\hat{f}$ is evaluated over $\mathcal{D}_{\text{test}}$, which is a separate dataset and assumed to be exclusive from the training dataset, i.e. $\mathcal{D}_{\mathrm{train}} \cap \mathcal{D}_{\mathrm{test}} = \emptyset$.

    \subsection{Data Augmentation} \label{sec:data_augmentation}
        \ac{DA} is a method widely used in \ac{DL} applications to improve generalizability and reduce overfitting by generating new data with respect to the domain of interest\cite{aug_survey}. It significantly increases performance over the test sets, namely generalization performance, and reduces the data collection burden. The methods employ a transformation operator $\mathcal{T}: \mathcal{X}^n \rightarrow \mathcal{X}$ applied over a (group of) sample point(s) from the training dataset and add the resulting sample to the training set, where $n$ is the number of samples used to generate new data by the transformation operator $\mathcal{T}$. As an example, we may consider a noise injection operator that takes a sample of complex channel frequency response $\boldsymbol{x}$ and outputs $\boldsymbol{x} + \boldsymbol{w}$ where $\boldsymbol{w} \sim \mathcal{CN} (0, \boldsymbol{I})$. 

In most of the applications, the corresponding label of the data $\boldsymbol{y}$ is kept the same, i.e., $f(\boldsymbol{x}) = f(\mathcal{T}(\boldsymbol{x}))$, where $f$ is the underlying true mapping between feature and label space. In regression tasks, labels may need further transformation. For image classification problems, a feature is the pixels of an image consisting of RGB channels. Then, common transformations applied to the sample are shifting colors, grays-cale transform, rotation, and zoom\cite{aug_survey}. Such transformations leverage the knowledge that images belong to the same class, whether they are transformed or not. This way, human knowledge about the different representations of a particular class is injected into the \ac{DL} algorithm by changing its input, the training dataset. 
        
\section{CSI-Based Indoor Localization} \label{sec:csi_indoor}
    \subsection{System Model} \label{sec:model}
         This section describes the transceiver and channel models used in this work. Assume that there are $N_{\mathrm{AP}}$ wireless \acp{AP}, and each \ac{AP} has $N_{\mathrm{RX}}$ antennas. The \ac{UE} is assumed to have a single antenna only. The system employs \ac{OFDM} with $M$ subcarriers. Without loss of generality, we assume that the localization is based on uplink transmission. Let $r_{j, k}$ be the received signal at the $j^{\text{th}}$ AP's $k^{\text{th}}$ antenna, where $j \in \{1,2, \dots N_{\mathrm{AP}}\}$, $k \in \{1,2, \dots N_{\mathrm{RX}}\}$. Further, let the transmitted signal from position $i$ at subcarrier frequency $f_m$ be $s_i(f_m)$, where $i \in \{1,2, \dots N\}$, and $m \in \{1,2, \dots, M\}$, and $H_{i, j, k}(f_m)$ be the channel frequency response at the $m^{\text{th}}$ subcarrier with respect to $j^{\text{th}}$ \ac{AP}'s $k^{\text{th}}$ antenna. Then, the received signal is given as follows:
\begin{equation}
    r_{i, j,k}(f_m) =  H_{i, j, k}(f_m)s_i(f_m) + w_{j,k}(f_m)
\end{equation}
where the noise samples $w_{j,k}(f_m) \sim \mathcal{CN}(0, \sigma^2_{\rm w})$ are i.i.d. zero-mean circularly symmetric complex Gaussian samples with variance $\sigma_{\rm w}^2$. 

The channel is assumed to be a wideband channel with $L$ \acp{MPC}. Then, the corresponding channel frequency response is 
\begin{equation}
    H_{i, j, k}(f_m)  \triangleq \sum_{l = 1}^L \alpha_l a_{j, k}(\phi_l, \theta_l, f_m)e^{-j2\pi f_m\tau_l}
\end{equation}
The channel impulse response is modeled as follows:
\begin{equation}
    h_{i, j,k}(\tau) \triangleq \sum_{l = 1}^L   \alpha_l a_{j,k}(\phi_l, \theta_l)\delta(\tau - \tau_l)
\end{equation}
where $a_{j,k}(\phi_l, \theta_l, f_m)$ is the antenna pattern of the $k^{\rm th}$ antenna element of the AP with respect to azimuth angle $\phi_l$, elevation angle $\theta_l$, $\tau_l$ is the delay of the $l^{\text{th}}$ component, and the \ac{MPC} has a complex amplitude gain $\alpha_l$; to simplify the model we assume here an isotropic antenna at the \ac{UE}. We generally assume the existence of a large number of \acp{MPC}, such that ''many" \acp{MPC} fall within each resolvable delay bin of width $1/B$, where $B$ is the system bandwidth. This leads to fading of $ H_{i, j, k}(f_m) $. As is common in the literature, we assume that this fading gives rise to a Rayleigh distribution (more precisely, a circularly symmetric zero-mean complex Gaussian distribution of the complex amplitude) within each bin, except for a delay bin containing a \ac{LOS} contribution \cite[Chap. 5-7]{prof_molisch}. We also assume that the \ac{WSSUS} is valid within a spatial region (whose size might depend on the environment). Note that the assumptions about fading inspire some of our proposed methods, but our evaluations of the efficacy of the algorithms in Sec. VI uses raw measured data; thus {\em do not depend} at all on these assumptions. 

Throughout the paper, we denote the $\boldsymbol{H}_{i,j,k}$ as a vector that contains all the channel frequency responses for the $ i^{\text{th}}$ sample's $j^{\text{th}}$ \ac{AP}'s $k^{\text{th}}$ antenna, similarly $\boldsymbol{h}_{i,j,k}$ is the vector that contains the channel impulses responses for all delay bins respectively. Note that we neglect the correlation between antenna elements since the antennas are spaced half-wavelength apart, and the angular spectrum typically shows a large spread\cite[Chapter 12]{prof_molisch}. Furthermore, different antennas usually have different patterns.
        
    \subsection{DL-Based Indoor Localization} \label{sec:dl_indoor}
        Here, we describe the features, labels, and loss functions used in the \ac{DL} algorithms for indoor localization, particularly the models used in this paper. For each measurement point $i$ in the environment, we have a total of $N_{\mathrm{AP}} \times N_{\mathrm{RX}}$ measurements, where each measurement contains responses from $M$ subcarriers. Since the feed-forward fully connected neural networks take real inputs as vectors, we first vectorize the measurements by concatenating the $N_{\mathrm{AP}} \times N_{\mathrm{RX}}$ measurement of each location. Then, the resulting complex vector is split into real and imaginary parts and concatenated together, resulting in a vector $\boldsymbol{H}_i$. As a result, the input feature is $\boldsymbol{x} \in \mathbb{R}^D$, where $D = M \times N_{\mathrm{AP}} \times N_{\mathrm{RX}} \times 2$.

On the other hand, CNNs take tensors as an input. We first create complex tensors in the shape of $N_{\mathrm{AP}}N_{\mathrm{RX}} \times M$. Then, we split real and imaginary parts of the CSI as channels of CNN input. Thus, a feature tensor for CNN is $\boldsymbol{x} \in \mathbb{R}^{N_{\mathrm{RX}}N_{\mathrm{AP}} \times M \times 2}$ We employ 2-D locations as the label, thus $\boldsymbol{y} \in \mathbb{R}^2$, which makes the output size of the employed neural nets 2. Finally, the resulting $N$ features and corresponding labels are coupled and used as the dataset $\mathcal{D} = \{(\boldsymbol{x}_i, \boldsymbol{y}_i)\}^{N}_{i = 1}$. 

The objective of the \ac{DL} algorithms is the minimization of the \ac{MSE} between a given location prediction and ground truth, which means that $\ell(\hat{\boldsymbol y}, \boldsymbol y) = \lVert \hat{\boldsymbol y} - \boldsymbol y  \rVert^2  = \sum_{q = 1}^2 (\hat{y}_q - y_q)^2$, where the subscript indicates the position within the vector of length two and $\hat{\boldsymbol{y}}$ is the corresponding prediction made by the model $\hat{f}$. 
The training objective is as follows: 
\begin{equation}
    \hat{f} = \argmin_{f\in \mathcal{F}} \frac{1}{N}\sum_{i = 1}^N \lVert f(\boldsymbol{x}_i) - \boldsymbol{y}_i \rVert^2
\end{equation}
where $\boldsymbol{y}_i$ is true 2-D location for the point $i$. 

We evaluate the performance of the \ac{DA} methods in terms of \ac{RMSE} of the trained models' results over the test dataset $\mathcal{D}_{\mathrm{test}}$, thus having units of meters. \ac{RMSE} of a model $f$ over a dataset $\mathcal{D}$ is: 
\begin{equation}
    \text{RMSE}(f, \mathcal{D}) = \sqrt{\frac{1}{N}\sum_{i = 1}^N \lVert f(\boldsymbol{x}_i) - \boldsymbol{y}_i \rVert^2}
\end{equation}
Note that indoor localization problems may alternatively entail a fingerprint or region classification problem using other feature and label vectors, as discussed in Sec. I.  
        
    \subsection{Problem Formulation} \label{sec:problem}
        Until now, we have shown how to pose the indoor localization problem with \ac{DL} algorithms and \ac{CSI} measurements. Here we formulate the \ac{DA} problem of interest. Assume that we are given a dataset $\mathcal{D}$ consists of \ac{CSI} measurements and position labels with a certain \ac{DL} algorithm $\mathcal{A}$ which maps from $\mathcal{D}$ to a model $f \in \mathcal{F}$, where $f: \mathcal{X} \rightarrow \mathcal{Y}$. Then, we would like to apply an augmentation operator $\mathcal{T}: \mathcal{X} \rightarrow \mathcal{X}$ such that the true risk of the output model is lower than the initial model, namely $R(f^{\star}) \leq  R(f)$. Here $R(f^{\star})$ is the true risk of the model $f^{\star}$, which is the output of the algorithm $\mathcal{A}$ fed by the augmented dataset $\mathcal{D}^{\star}$. Note that we do not have access to the true data distribution $P_{\mathcal{X}\mathcal{Y}}$, then we estimate the true risk by empirical risk over the test dataset $\mathcal{D}_{\mathrm{test}}$. 
        
\section{Data Augmentation Algorithms Based on Transceiver Behavior} \label{sec:transceiver} 

    \subsection{Random Phase} \label{sec:random_phase}
        During the measurement of the channel responses, local oscillators are employed in the up/down converters at the \acp{AP} as well as the \ac{UE}. Since no oscillator is ideal, there are phase drifts of the AP oscillators with respect to the \ac{UE} oscillator. It is reasonable to assume that the \acp{AP}' clocks drift independently from each other, and the phase state at a given time can be modeled as independent random variables uniformly distributed between $0$ and $2\pi$. Since they are all derived from the same local oscillator, the phase offsets are assumed to be the same for all measurements made at the different antenna elements and subcarriers. Algorithm \ref{algo:phase_via_ap} below summarizes the process. The algorithm takes $N$ pairs, each consisting of measurements made by all \acp{AP} at measurement location $i$, i.e., $\boldsymbol{H}_i$, and the corresponding labels $\boldsymbol{y}_i$ for the measurement point $i$. Then, for each sample and each \ac{AP}, a random phase is generated and added to all measurements. The labels are kept identical, and we call this algorithm \texttt{PHASE\_AP}.
\begin{algorithm}[h!]

        \KwIn{$\mathcal{S}$ = $\{(\boldsymbol{H}_i, \boldsymbol{y}_i)\}_{i = 1}^N$, $N^{\star}$}
        \KwOut{$\mathcal{S}^{\star} = \{(\boldsymbol{H}_i, \boldsymbol{y}_i)\}_{i = 1}^{N^{\star}}$}
        $\mathcal{S}^{\star} \gets \mathcal{S}$\Comment{Initialization} \\
        \While{$ |\mathcal{S}^{\star}|\leq N^{\star}$}{
            \For {\text{each sample} $i$}{
                \For {\text{each AP} $j$}{
                    \uIf{\texttt{PHASE\_AP}}{
                  $\phi \gets \mathcal{U}[0,2\pi]$ \Comment{Phase Shift}\\
                  ${\boldsymbol{H}_{i, j}^{\star}} \gets \boldsymbol{H}_{i, j} \times e^{j\boldsymbol{\phi}}$ \Comment{Add shift}\\
                    }
                    \uElseIf{\texttt{AMP\_AP}}{
                        $\alpha \gets \mathcal{U}[-P^{\star},P^{\star}]$ \Comment{Amplitude Shift} \\
                        ${\boldsymbol{H}_{i, j}^{\star}} \gets \boldsymbol{H}_{i, j} \times 10^{\alpha/10}$ \Comment{Add shift}
                    }
                }
            $\mathcal{S}^{\star} \gets \mathcal{S}^{\star} \cup ({\boldsymbol{H}^{\star}_i}, \boldsymbol{y}_{i})$ \Comment{Add new samples}\\
            }    
        }
        \caption{Transceiver Based Data Augmentations via AP}
        \label{algo:phase_via_ap}
\end{algorithm}
A variant of this algorithm is to independently add phase drift to each antenna on a given AP, corresponding, e.g., to each antenna having an independent oscillator or at least an independent phase-locked loop that might introduce phase noise. The corresponding changes to the algorithm \ref{algo:phase_via_ap} are straightforward: we add an inner loop for each RX antenna $k$ on AP $j$. Then, we generate: $\bold{\phi} \gets \mathcal{U}[0,2\pi]$, and apply the new phase to the measured channel frequency responses as follows: $\boldsymbol{H}_{i, j, k}^{\star} \gets \boldsymbol{H}_{i, j, k} \times e^{j\boldsymbol{\phi}}$. We refer to this variation as \texttt{PHASE\_RX}.

    \subsection{Random Amplitude} \label{sec:random_amp}
        Real-world hardware leads not only to fluctuations of the phase but also of the amplitude due to variations of the amplifier gains in the low noise amplifier and automatic gain control in the receivers in the APs. We model 
such fluctuations as random variables, which are uniformly generated over a pre-defined interval, i.e., $[-P^{\star}, P^{\star}]$ dB independently for each AP. Then, this amplitude is added (on a dB scale) to all signals from that measurement location, similar to the procedure in Algorithm \ref{algo:phase_via_ap}. 
In addition to this uniform distribution, we also studied further different distributions and variances, including Gaussian, Laplace, and Gaussian mixtures with 0.1, 0.5, 1, and 1.5 dB variances, where distributions are zero-mean except for the mixture model, which consists of two Gaussians which are centered around $-P^{\star}$ and $P^{\star}$. However, since the alternative distributions did not perform significantly better than the uniform distribution, we employ uniform distributions in the results shown in Sec. VI. Algorithm \ref{algo:phase_via_ap} provides a detailed procedure description, where this approach is called \texttt{AMP\_AP}.

The algorithm can be extended to independently add amplitude drift to each antenna on a given AP,  similar to the previous method, which is called \texttt{AMP\_RX}. Note that this approach emulates amplitude variations in the transceiver, not fading of the channel; it is also different from random noise injection. 

\section{Data Augmentation Algorithms Based on Channel Behavior} \label{sec:channel_algos}
    \subsection{Data Augmentation via Correlation} \label{sec:corr_algo}
        This algorithm is based on creating different realizations of the small-scale fading consistent with a measured transfer function. We create channel realizations that will occur somewhere in the close vicinity (typically $10-40 \lambda$) of the measurement point and assign them the same location label as the measured point. The creation of the different small-scale realizations is based on the \ac{WSSUS} assumption and the fact that frequency and spatial variations of the channel gain are caused by the same physical effect, namely the complex superposition of the different MPCs \cite{prof_molisch}. We drop the subscripts and use $\boldsymbol{H} \in \mathbb{C}^M$, i.e. $\boldsymbol{H} = [H_1, H_2, \dots H_M]$ as frequency response vector for measurement for a point $i$, AP $j$, and RX $k$. Here, $H_m$ is the channel frequency response for frequency $f_m$. Then, the normalized frequency correlation matrix is:
\begin{equation}
    \boldsymbol{\Sigma} = \frac{\mathbb{E}_{f}[\boldsymbol{H}\boldsymbol{H^{\dagger}}]}{\frac{1}{M}\sum_{m = 1}^M\mathbb{E}[H_mH_m^{\dagger}]}
\end{equation}
where $m \in \{1, \dots, M\}$ and $\dagger$ corresponds to Hermitian transpose. Then, by the US assumption: 
\begin{align}
    \mathbb{E}_f[\boldsymbol{H}\boldsymbol{H^{\dagger}}] &= \begin{bmatrix}R(0)  & \ldots  & R(M-1) \\ \vdots & \ddots & \vdots \\ R(1 - M) & \ldots & R(0)\end{bmatrix}
\end{align}
where $R(\Delta) \triangleq \mathbb{E}_m[H_{m}H_{m+\Delta}^{\dagger}]$ is the \ac{ACF} in the frequency domain. Since observation of the ensemble is not available, we approximate the expectation $R(\Delta) \approx \hat{R}(\Delta)$ as  follows (using the measurements we have):
\begin{equation}
    \hat{R}(\Delta)  = \begin{cases} \frac{1}{|\mathcal{V}_{\Delta}|}\sum_{i, j \in \mathcal{V}_{\Delta}} H_iH_j^{\dagger}, &\text{if } \Delta \leq \Delta^*  \\ 0, & \text{otherwise} \end{cases}
\end{equation}
where $\mathcal{V}_{\Delta}$ is the set of indices $|i - j| = \Delta$, $i < j$ and $\Delta \in \{0, 1 \dots (M - 1)\}$. Setting $\hat{R}(\Delta) = 0$ for $\Delta > \Delta^*$ is physically meaningful in most wideband systems, more specifically if the system bandwidth is much larger than the coherence bandwidth $B_{\rm coh}$ of the channel (implying that $\Delta^*>>B_{\rm coh}$ should be fulfilled). It is also mathematically desirable to avoid numerical instabilities when the ACF is estimated from a small number of empirical values, i.e., $|\mathcal{V}_{\Delta}|$ is small. $\Delta$ is selected empirically during hyperparameter tuning. As we assumed a WSSUS model, each realization of ${H}_m$ obeys the above correlation estimations. Making furthermore the assumption of zero-mean complex Gaussian fading, we can create transfer functions as realizations of $\boldsymbol{H} \sim \mathcal{CN}(0, \boldsymbol{\Sigma})$ where $\boldsymbol{\Sigma}$ is the covariance matrix. With this assumption, we expect this method to work better in \ac{NLOS} environments than \ac{LOS} environments.

To draw new realizations, an uncorrelated complex Gaussian random vector $\boldsymbol{x} \sim \mathcal{CN}(0, \hat{R}(0))$ is generated. Then, the new correlated channel responses are generated as follows: $\boldsymbol{H}^{\star}_{i,j,k} \sim \boldsymbol{C}\boldsymbol{x}$ where $\hat{\boldsymbol{\Sigma}} = \boldsymbol{C}\boldsymbol{C}^{\dagger}$. This decomposition can be made with a choice of a generic algorithm such as Cholesky. However, the {\em estimated} correlation matrix $\hat{\boldsymbol{\Sigma}}$ is not always positive definite; hence, we cannot use such decomposition. There are quite a few methods to find the closest positive definite matrices, but - to the best of our knowledge - not for complex matrices. Thus, we {\em empirically} take the matrix square root $\hat{\boldsymbol{\Sigma}} = \boldsymbol{C}\boldsymbol{C}$; we verified - for the channels in Sec. VI - that the normalized error of $\lVert\hat{\boldsymbol{\Sigma}} - \boldsymbol{C}\boldsymbol{C}^{\dagger}\rVert_{F}/\lVert \boldsymbol{C}\boldsymbol{C}^{\dagger}\rVert_{F}$ is very small, i.e. 0.05, where $\lVert \cdot \rVert_{F}$ Frobenius norm of a given matrix. Thus, we can create $\boldsymbol{H} = \boldsymbol{C}\boldsymbol{x}$ as correlated random variables. The algorithm is summarized in Algorithm \ref{algo:correlation}, where we call this approach \texttt{CORR}.
\begin{algorithm}[h!]
        \KwIn{$(\boldsymbol{H}_{i,j,k}, \boldsymbol{y}_i), \hat{\boldsymbol{\Sigma}}$}
        \KwOut{$(\boldsymbol{H}_{i,j,k}^{\star}, \boldsymbol{y}_i))$}
        \For{each $\Delta \in \{0, \dots, M - 1\}$}{
            \uIf{$\Delta \leq \Delta^{\star}$}{
                $\hat{R}(\Delta)$ $\gets$ $\frac{1}{|\mathcal{V}_{\Delta}|}\sum_{m, n \in \mathcal{V}_{\Delta}} H_{i,j,k}(f_m)H_{i,j,k}^{\dagger}(f_n)$
            }
            \Else{
                $\hat{R}(\Delta) \gets 0$ \Comment{All zeroes after certain $\Delta^{\star}$}
            }
            \uIf{$ \Delta = m - n $}{
                $\hat{\boldsymbol{\Sigma}}_{mn} \gets \hat{R}(\Delta)/\hat{R}(0)$ \Comment{Assign Normalized Autocor.}
            }
            \uElseIf{$\Delta = n- m$}{
                $\hat{\boldsymbol{\Sigma}}_{mn} \gets \hat{R}^{\dagger}(\Delta)/\hat{R}(0)$ \Comment{Assign conjugate}
            }
        }
        $\boldsymbol{C} \gets \hat{\boldsymbol{\Sigma}}^{1/2}$ \\
        $\boldsymbol{x} \sim \mathcal{CN}(0, R(0))$\\
        $\boldsymbol H_{i, j, k}^{\star} \gets \boldsymbol{C}\boldsymbol{x}$ \Comment{New Sample}
        \caption{\texttt{CORR}:Augmentation via Correlation}
        \label{algo:correlation}
\end{algorithm}
where $\hat{\boldsymbol{\Sigma}}_{mn}$ is the element placed in $m^{\text{th}}$ row and $n^{\text{th}}$ column. Algorithm \ref{algo:correlation} is repeated for all the sample points and corresponding AP and RX for the measurements individually. When multiple measurements within a stationarity region are available, these can also be used to improve the estimates of the correlation matrix.

    \subsection{PDP Based Data Augmentation Methods} \label{sec:pdp_algos}
        \subsubsection{PDP 1, Random Phase over Delay Bins} \label{sec:pdp1}
Another way to generate different channel realizations is to set the magnitude of the impulse response equal to the square root of the PDP (i.e., leave the magnitude of the original measurement unchanged) while generating different random phases; these phases are independent between delay bins, and independent between different channel realizations. This approach is justified in the case where each delay bin has only one MPC so that a small change of the UE location results only in a phase change; for rich multipath, it is meaningful for the bin containing the LOS component (if such exists). 

Algorithm \ref{algo:pdp124} summarizes the procedure for one measurement location, where we call this approach \texttt{PDP1}. We first take the Inverse Fast Fourier Transform(IFFT) of the measured channel frequency responses; then, we take the amplitude of each delay bin and generate new random impulse responses via randomly generated phases for each bin. This procedure is repeated for all measurement points and corresponding APs and RXs. The labels for the generated samples are the same as the original sample labels. 

        \subsubsection{PDP 2, Rayleigh Fading} \label{sec:pdp2}
For the case where we have many MPCs in each delay bin, movement of the UE will result in a change of both magnitude and phase. We conjecture that such a method is more helpful in Non-Line-of-Sight(NLOS). The impulse response is generated as a zero-mean complex Gaussian variable with a variance corresponding to the PDP value in that bin. Subsequently, we convert the impulse responses back to the frequency domain via FFT. We provide the procedure for a single measurement vector for a particular point, AP, and RX in Algorithm \ref{algo:pdp124} under the name of \texttt{PDP2}.

        \subsubsection{PDP 3, Averaging over Cell} \label{sec:pdp3}
            \begin{algorithm}[h!]
    \KwIn{$\{(\boldsymbol{H}_{c_i}, \boldsymbol{y}_{\mathcal{C}})\}_{c_i \in \mathcal{C}}$}
    \KwOut{$\{(\boldsymbol{H}^{\star}_{c_i}, \boldsymbol{y}_\mathcal{C})\}{c_i \in \mathcal{C}}$}
    \For {\text{each sample $c_i$ in cell} $\mathcal{C}$}{
        \For{\text{each AP} $j$}{
            \For{\text{each RX} $k$}{
                $\boldsymbol{h}_{c_i,j,k} \gets \text{IFFT}(\boldsymbol{H}_{c_i,j,k})$ \Comment{Delay Domain}\\
        }
            \For{\text{each RX} $k$}{
                \For{\text{each delay bin} $m$}{
                    $P_c \gets \frac{1}{|\mathcal{C}|}\sum_{c_i \in \mathcal{C}} |h_{c_i,j,k}(\tau_m)|^2$ 
                    ${h}_{j, k}^{\star}(\tau_m) \sim \mathcal{CN}(0,P_c)$ \Comment{Avg. Power}\\ 
                    }
                    $\boldsymbol{H}_{c_i, j,k}^{\star} \gets \text{FFT}(\boldsymbol{h}^{*}_{c_i, j,k})$ \Comment{New Freq. Resp.}\\
               }
            }   
    }
    \caption{PDP Based Augmentation 3 with Rayleigh Fading over Cells}
    \label{algo:pdp3}
\end{algorithm}
Previously, in Algorithm \texttt{PDP2}, we generated Rayleigh fading with respect to the measured PDP. However, a single realization as the base PDP of newly generated samples could lead to biased results. Thus, we propose creating uniformly spaced areas (called cells henceforth) over the environment and averaging the PDPs over all of the measurements in the cell. We use this averaged PDP to generate new channel impulse responses according to the method of PDP 2.

Algorithm \ref{algo:pdp3} takes cell centers as labels and newly generated samples have the same label $\boldsymbol{y}_{\mathcal{C}}$. Other labeling schemes are also considered, such as generating labels uniformly over the cell or using the same labels as the original measurements. However, during experiments, we see using cell centers as labels provided better localization performance when this DA method was applied. We also experimented with different cell spacing, i.e., 0.5m, 1m, 1.5m and 2m. We found that 1m spacing yielded better test set results. 

An important caveat for this algorithm is that it trades off the better trainability and increased robustness created by the \ac{DA} with the spatial resolution. Specifically, we assign all channel realizations (which represent a {\em region} of stationarity, typically of size $10-20\lambda$), a {\em single} location label. This causes a loss in spatial resolution. The algorithm is thus mainly beneficial in situations where only a small number of training data is available since, in that case, the benefits of augmentation outweigh the loss of resolution. The algorithm is summarized in \ref{algo:pdp3} for the samples in the cell $\mathcal{C}$. 
            
        \subsubsection{PDP 4, A Mixed Approach} \label{sec:pdp4}
            This method will follow a mixed approach, using both random phase injection to the \ac{PDP} and Rayleigh fading channel. We create new realizations for the Rayleigh fading channel, which primarily models the \ac{NLOS} channels, except for the highest power component in the delay domain. For the highest delay bin, we impose a random phase on the measured magnitude. We conjecture that this augmentation method might be more suitable for \ac{LOS} channels. The labels for the generated samples are kept the same as those for the original samples. We summarize the procedure for a single sample point $i$ and its specific AP $j$ and RX $k$ in Algorithm \ref{algo:pdp124}, in the case of \texttt{PDP4}. We also note that further refinements could be done by detecting whether a particular link has \ac{LOS} or not - a problem for which a variety of methods exist in the literature. However, investigation of those methods is beyond the scope of the current paper. 
\begin{algorithm}[h!]
        \KwIn{$(\boldsymbol{H}_{i,j,k}, \boldsymbol{y}_i)$}
        \KwOut{$(\boldsymbol{H}_{i,j,k}^{\star}, \boldsymbol{y}_i)$}
        $\boldsymbol{h}_{i,j,k} \gets \text{IFFT}(\boldsymbol{H}_{i,j,k})$ \Comment{CIR}\\
        \For{\text{each delay bin} $m$}{
            \uIf{\texttt{PDP1}}{

                $\phi \sim \mathcal{U}[0, 2\pi]$ \Comment{New Phase}\\    
                $h_{i,j,k}^{\star}(\tau_m) \gets |h_{i,j, k}(\tau_m)|\times e^{j\phi}$
            }
            \uElseIf{\texttt{PDP2}}{
                        $h_{i, j,k}^{\star}(\tau_m) \sim \mathcal{CN}(0,|h_{i,j, k}(\tau_m)|^2)$ \\ 
            }

            \uElseIf{\texttt{PDP4}}{
                \uIf{$\tau_m \neq \argmax_{\tau} |h_{i,j,k}(\tau)|$}{
                    $h_{i,j,k}^{\star}(\tau_m) \sim \mathcal{CN}(0,|h_{i,j,k}(\tau_m)|^2)$ \\
                }
                \Else{
                    $\phi \sim \mathcal{U}[0, 2\pi]$ \Comment{New Phase}\\    
                    $h_{i, j,k}^{\star}(\tau_m) \gets |h_{i,j, k}(\tau_m)|\times e^{j\phi}$ \\
                }
            }
        }
        $\boldsymbol{H}_{i,j,k}^{\star} \gets \text{FFT}(\boldsymbol{h}_{i, j,k}^{\star})$ \Comment{New Freq. Resp.}
        \caption{PDP Based Augmentations 1-2-4}
        \label{algo:pdp124}
\end{algorithm}

\section{Numerical Evaluation} \label{sec:num_results}
    \subsection{Datasets} \label{sec:datasets}
        
We evaluate the methods and scenarios in four real datasets, WILD 1 Env. 1, WILD 1 Env. 2, WILD 2 Env. 1 and WILD 2 Env. 2., introduced in \cite{wild1, wild-v2}, respectively. Each dataset is based on measurements in a different environment, where WILD 1 datasets are mostly LOS, while Env. 2 includes an NLOS AP. Besides, the environments in WILD 2 datasets are highly NLOS and consist of long hallways, and many scatterers can be found in office setups. 
\begin{table}[h!] 
        \begin{tabular} {l l l l l}
        	\toprule[1pt]
        	\textit{\textbf{Params.}} & \textit{\textbf{WILD1-1}} & \textit{\textbf{WILD1-2}} & \textit{\textbf{WILD2-1}} & \textit{\textbf{WILD2-2}} \\
        	\cmidrule(lr){1-1} \cmidrule(lr){2-2} \cmidrule(lr){3-3} \cmidrule(lr){4-4} \cmidrule(lr){5-5}
        	Area & $46$ & $139$  & $400$ & $418$\\
            $BW$ &  80 MHz &80 MHz  & 80 MHz & 80 MHz   \\ 
            $f_c$ & 5 GHz& 5 GHz& 5 GHz & 5 GHz \\  
        	$N_{\mathrm{AP}}$ &   3 &   4 & 6   & 6 \\  
            $N_{\mathrm{RX}}$ &   4 & 4& 4& 4\\  
            $N$ & 56395 & 51613 & 17000 & 5000\\
            $M$ & 234 & 234 & 234 & 234 \\
        	\bottomrule[1pt]
        \end{tabular}
    \caption{Dataset Details}
    \label{table:dataset}
\end{table}

The complex channel frequency response data collected for each sampling point is in the format $\boldsymbol{x} \in \mathbb{C}^{M \times N_{\mathrm{AP}} \times N_{\mathrm{RX}}}$. We separate each dataset into three sub-datasets, namely, training, validation, and test, which are mutually exclusive. We randomly selected 32000 samples for training in WILD 1 environments and 12000 and 3000 in WILD 2 environments, respectively. The rest are used for validation and testing. We used the WILD 1 dataset without any processing. However, the WILD 2 dataset contains padding elements, i.e., zero-valued measurements in the dataset. We eliminate any zero-valued entry before using it in the simulations. More detailed dataset information is presented in Table \ref{table:dataset}, where Area is given in terms of square meters.

    \subsection{Neural Network Architectures} \label{sec:neural_nets_arch}
        To demonstrate the localization performance of the proposed DA methods, we employ two different architectures, \ac{FCNN} and \ac{CNN} networks. When picking such architecture types, the primary consideration is the training and evaluation (validation and test) time. In \cite{csi_attention}, it is shown that advanced modules such as attention-based mechanisms are state-of-the-art neural network-based localization modules, yet they needed a longer time to train such cases. Since our experimentation includes repetitive training and testing due to different dataset types, sizes, augmentation ratios, methods, and scenarios, we needed to reduce the time spent. Thus, we focused on simple yet still well-performing architectures \cite{hoydis}. 

These architectures employ different hyperparameters. We ran a grid-wise hyperparameter search by training each configuration and comparing and picking the best hyperparameters with respect to the best validation loss. For \ac{FCNN}, we searched the following parameters: batch size, dropout probability, fully connected layer width, and number of fully connected hidden layers. For \ac{CNN}: number of channels, convolutional layers with average pooling, stride size, kernel size, fully connected layer width, and number of fully connected hidden layers. The resulting \ac{FCNN} is 4 hidden layers with the size of 512 neurons, with each layer followed by a 0.2 probability dropout.
On the other hand, the final \ac{CNN} architecture is 3 2D convolutional layers of channel size 64 with kernel size $(1, 2)$ and stride $(1, 2)$, where an average pooling follows each. The convolutional structure is followed by 3 hidden, fully connected neural layers with a size of 512, where each layer is applied dropout with a probability of 0.2. Moreover, we used the activation function ReLU, i.e., $\phi(x) = \max(0, x)$. 

We trained the models for 50-75 epochs with $10^{-4}$ learning rate, $10^{-5}$ weight decay, and batch size 32. We used a batch size of 32. The training took place on either NVIDIA A100 or NVIDIA A5000 GPUs.

    \subsection{Main Results} \label{sec:main_results}
        This section illustrates the effect of the \ac{DA}. For all figures, we run the algorithms in five independent trials and present the average performance as a line/curve and the variations around the average as a shaded region. The x-axis denotes the augmentation factor, i.e., how many augmentation samples are created for each measured sample. Thus, the first points in the plots (augmentation factor 0)  show the raw performance, i.e., without augmentation. The y-axis is the test set MSE in units of meters. The phrase ''original dataset size" refers to the number of actual measurements in the dataset. For example, if the original dataset consists of 100 measurements, then all the 100 measurements from the environment are used, and the rest of the data is generated via the proposed augmentation methods. In the simulations, we generally use subsets of the total available measured training data in order to explore the impact of different amounts of measured data. These subsets are chosen randomly before all the experiments. So, the experiments use the same training, validating, and testing splits to make the comparisons fair. Finally, we compare results based on \ac{DA} with results obtained with the full measured datasets (without augmentation labeled as \textit{Full}). The corresponding dataset sizes are 32000 for both of the WILD1 environments and 1200 and 3000 for WILD2 Env. 1 and WILD2 Env. 2, respectively.

        \subsubsection{Low Data Regime} \label{sec:low_data}
            
The collection of large amounts of data in the environment of interest may be costly or even infeasible in many cases. 
The very low data regimes, where there is access to only 100 to 1000 measurements taken from the environment, are thus of considerable interest. This section analyzes the performance of our \ac{DA} methods in two different environments, WILD1 Env. 1, \ac{LOS}, and WILD2 Env 2, \ac{NLOS}, where the training datasets include only 100 original measurements to emulate the low data regime case. 

\begin{figure}[h!]
    \centering
    \vspace{-5mm}
    \includegraphics[width =\columnwidth]{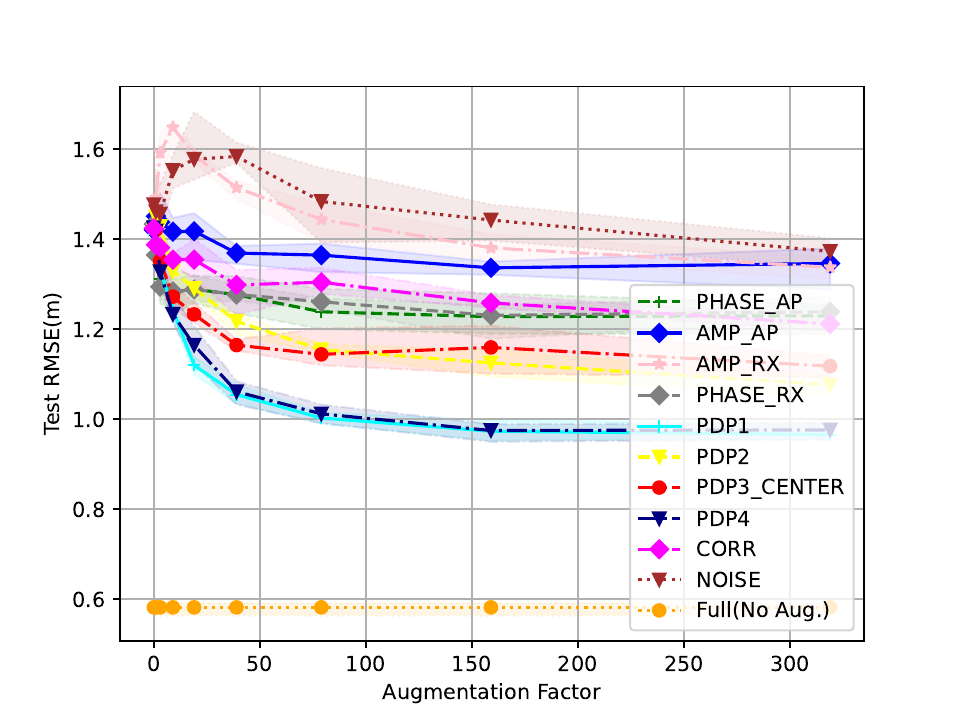}
    \caption{Test Set Performance vs Augmentation Size and Methods, Model: CNN, Dataset: WILD1 Env. 1, Original Dataset Size: 100}
    \label{fig:low_los}
\end{figure}
For the purpose of comparison with the state of the art, we also provide results with a simple noise injection method, where $\mathcal{T}(\boldsymbol{H}) = \boldsymbol{H} + \boldsymbol{n}$, where $\boldsymbol{n} \sim \mathcal{C}(0, P)$. Here, $P$ is the noise power level, and it is empirically tuned from a set of target Signal-to-Noise Ratios (SNRs) consisting of 0 to 20 dB. It is found that 20 dB and 15 dB SNRs are the best performing for \ac{LOS} and \ac{NLOS} environments, respectively. Results will show that noise injection performs significantly worse than (most of) our proposed augmentation methods; it will thus be omitted in later subsections. 

\begin{figure}[h!]
    \centering
\vspace{-5pt}
    \includegraphics[width =\columnwidth]{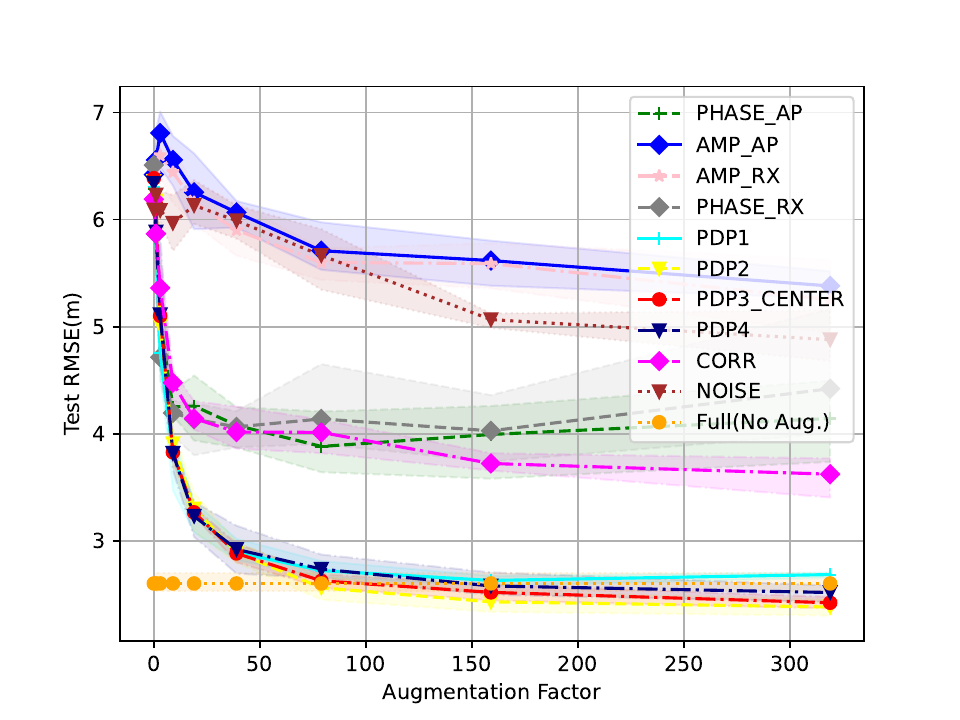}
    \caption{Test Set Performance vs Augmentation Size and Methods, Model: CNN, Dataset: WILD2 Env. 2, Original Dataset Size: 100}
    \label{fig:low_nlos}
\end{figure}

Figs. (\ref{fig:low_los}) and Fig. (\ref{fig:low_nlos}) show the results for low data regimes in a LOS and NLOS environment, respectively. All of the datasets are augmented up to 32000 samples. In the LOS scenario, performance improvement ranges from no improvement (AMP-AP method; the AMP-RX method may even reduce the accuracy) to improvements of about $30$\%. PDP 1 and PDP 4 methods perform best in this environment. This is intuitive since they both conserve the \ac{LOS} component's magnitude in the PDP. On the other hand, methods that implicitly assume Rayleigh fading (correlation, PDP 2, and PDP 3) perform worse. In no case does the augmentation method achieve the performance of the "full measured" dataset, with a gap of about $20$ \% remaining. 

This picture changes significantly in the NLOS environment. The RMSE is generally much larger there (in the un-augmented case, $6.5$ m versus $1.4$ m in the LOS case). Secondly, all methods lead to performance improvements, with the AMP-AP now performing worst ($10$ \% improvement), while PDP 2 and PDP 3 perform best (though PDP 1 and PDP 4 perform almost as well), with performance improvements of $50$ \%. Remarkably, performance with the augmentation methods can match the performance of a "full measurement set".

        \subsubsection{Medium Data Regime} \label{sec:med_data}
            To illustrate a case where the dataset is between low and high data regimes, we present localization performances of DA methods in a dataset with 8000 measurements, which is then augmented up to 96000 samples.
\begin{figure}[h!]
    \centering
    \includegraphics[width =\columnwidth]{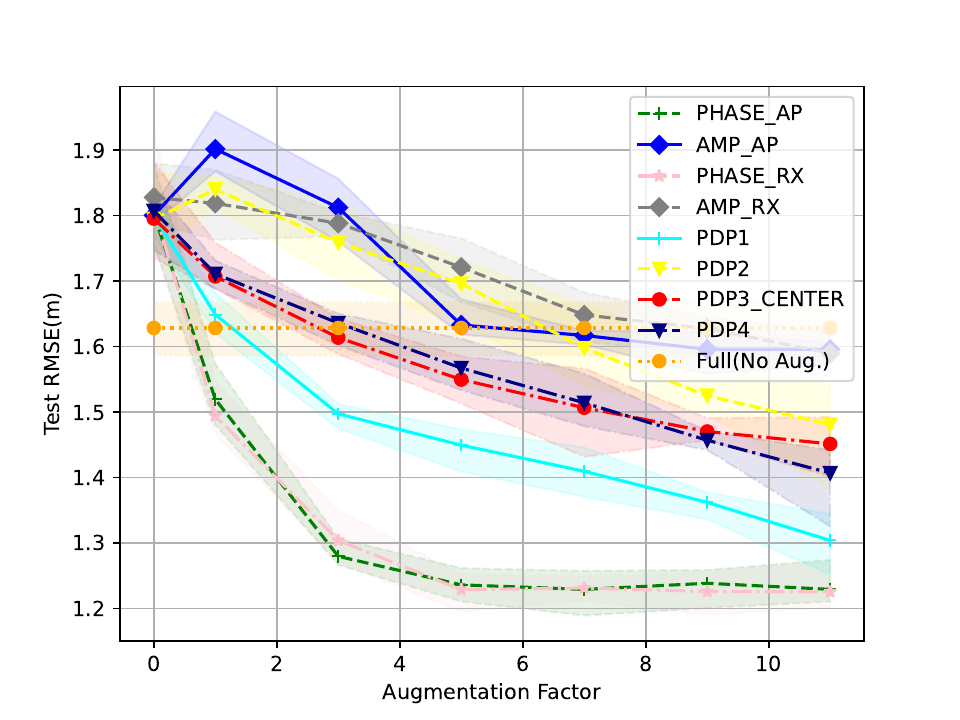}
    \caption{Test Set Performance vs Augmentation Size and Methods, Model: CNN, Dataset: WILD2 Env. 1, Original Dataset Size: 8000}
    \label{fig:med_8k_cnn_wild2_env1}
\end{figure}
Fig. (\ref{fig:med_8k_cnn_wild2_env1}) presents the localization accuracy in a highly NLOS environment. Adding random phases to the measurements brings a significant performance boost, with up to $66\%$ performance improvement. The phase augmentation (transceiver-based method) outperforms all channel-based methods. Remarkably, considering the NLOS environment, PDP2 (based on Rayleigh fading) performs worse than the other methods, indicating that assuming the powers of the delay bins from a single realization of the channel is not a good approximation; PDP3 (averaging over multiple points in the cell) performs significantly better. PDP4, which is adapted to the physics of both LOS and NLOS channels, performs comparable to PDP3. 
            
        \subsubsection{High Data Regime} \label{sec:high_data}

\begin{figure}[h!]
    \centering
    \vspace{-5mm}
    \includegraphics[width =\columnwidth]{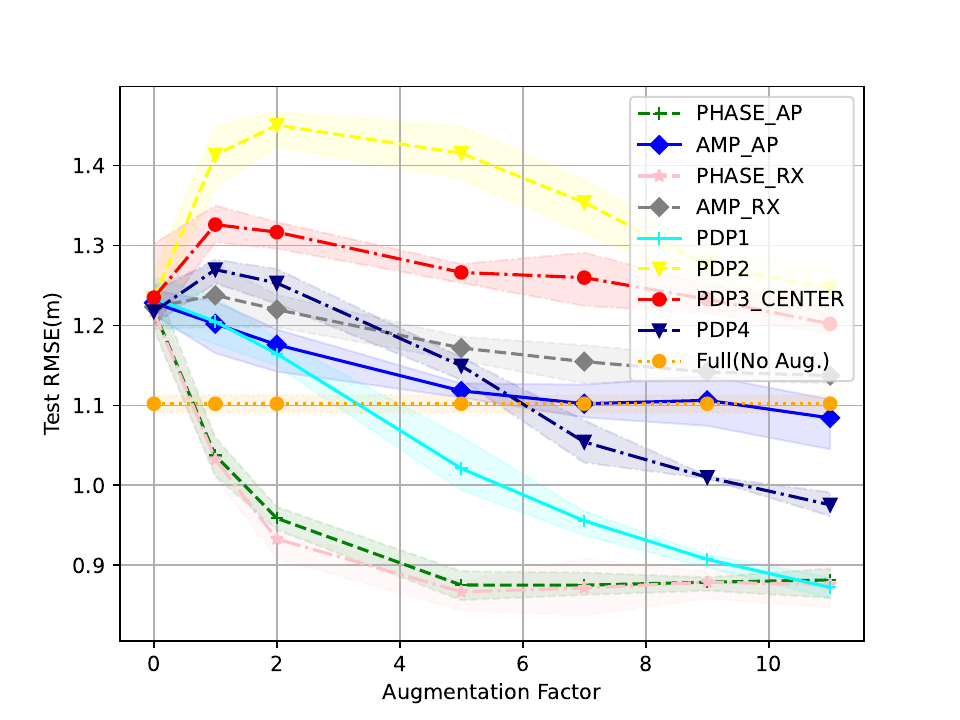}
    \caption{Test Set Performance vs Augmentation Size and Methods, Model: CNN, Dataset: WILD1 Env. 2, Original Dataset Size: 16000}
    \label{fig:high_16_cnn_wild1_nlos}
\end{figure}

To emulate the high-data-rate case, we conducted a simulation where the original dataset size is 16000 in a \ac{LOS} environment. Fig. (\ref{fig:high_16_cnn_wild1_nlos}) demonstrates that transceiver-based methods, particularly random phase methods, outperform other methods. 

We conjecture that as the dataset contains more measurements from the environment, it gets harder to meaningfully add even more channel realizations, which are furthermore based on assumptions. Compared to the low data regime in the LOS environment(see Fig. \ref{fig:low_los}), PDP 1 and PDP 4 methods perform worse yet do better than PDP 2 and PDP 3 methods. In a low data regime, Rayleigh fading-based augmentation did not hurt the localization performance, but imposing such a model in a high data regime actually decreased the localization accuracy. We augmented the dataset up to 192000 data points for all the simulations. Apart from the PDP 2 and PDP 3 methods, other augmentation methods outperform the full measurement-only training set (32000 measurements) without augmentation.   
            
        \subsubsection{Performance over Varying Number Measurements} \label{sec:num_measurements}
            
\begin{figure}[h!]
    \centering
    \vspace{-5mm}
    \includegraphics[width =\columnwidth]{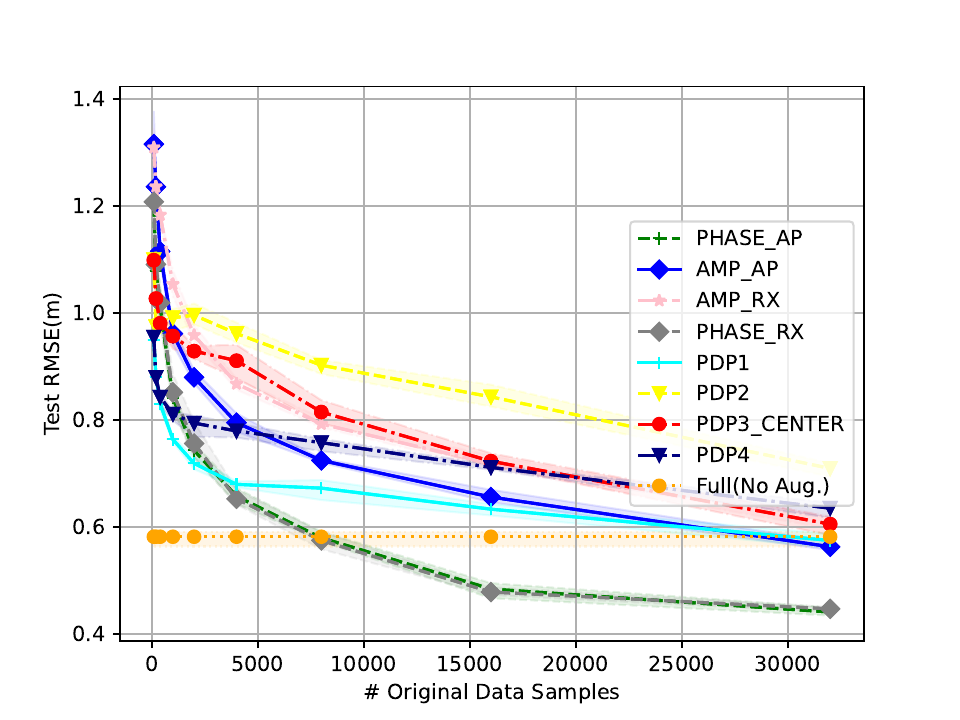}
    \caption{Test Set Performance vs Original Dataset Size and Aug. Methods, Model: CNN, Dataset: WILD1 Env. 1, All training sets augmented to the 64000 samples}
    \label{fig:original_size}
    \vspace{-3mm}
\end{figure}

The proposed DA methods were evaluated for different augmentation ratios in different data regimes. It is apparent that the number of measurements taken from the environment is an important aspect of localization accuracy. Fig. (\ref{fig:original_size}) presents the effect of the number of measurements. The x-axis denotes the number of measurements in the training dataset, and each case is augmented to 64000 training points. All methods' overall localization accuracy increased when the number of measurements increased. The transceiver-based methods outperform the channel model-based methods as the models learn the actual channel model when there are enough measurements in the training dataset. Imposing a channel model using the DA method is ineffective for high data regimes. 

        \subsubsection{Performance over Different Datasets} \label{sec:dataset_effect}
            
\begin{figure}[h!]
    \centering
    \vspace{-5mm}
    \includegraphics[width =\columnwidth]{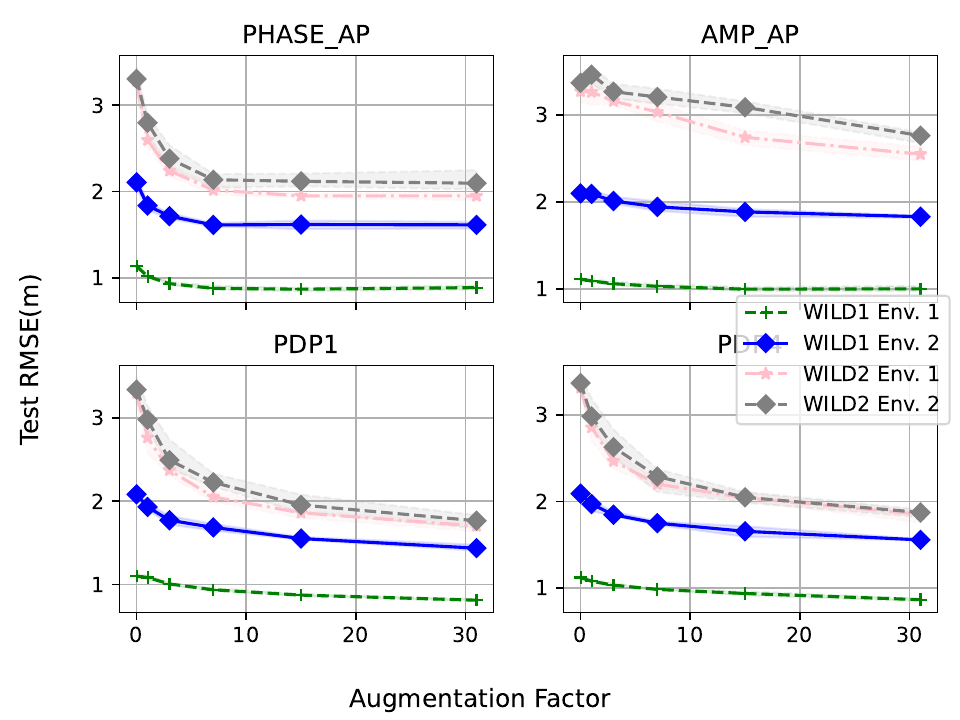}
    \caption{Test Set Performance vs Augmentation Size and Methods, Model: CNN, Original Dataset Size: 1000}
    \label{fig:dataset_low}
\end{figure}

\begin{figure}[h!]
    \centering
    \vspace{-5mm}
    \includegraphics[width =\columnwidth]{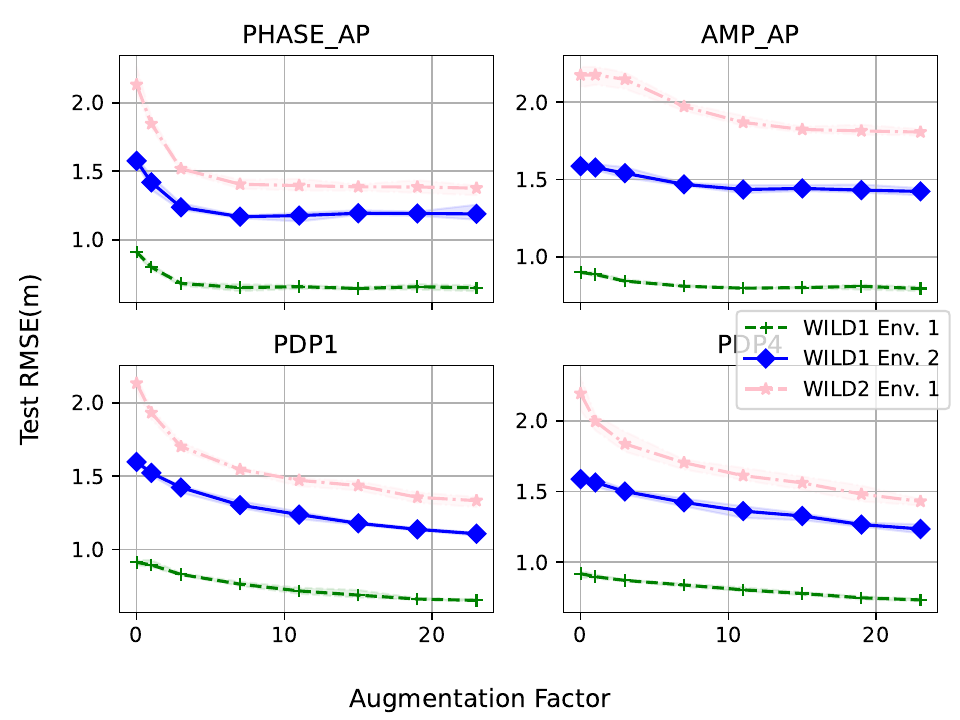}
    \caption{Test Set Performance vs Augmentation Size and Methods, Model: CNN, Original Dataset Size: 4000}
    \label{fig:dataset_med}
\end{figure}

The number of real measurements and the augmentation factor are two important aspects that have been evaluated previously to show how they affect the performance of DA methods. The environment/dataset plays a significant role as well. We employed a subset of methods in two different data regimes to highlight such a role. From Fig.(\ref{fig:dataset_low}) and Fig. (\ref{fig:dataset_med}), we observe that in a \ac{NLOS} (up to $43\%$) environment, we get more performance boost than in LOS (up to $25\%$).

Secondly, we must take into account the size of the area that the datasets cover. Augmentation helps to cover larger areas (WILD2 environments are significantly larger than WILD1) with less data and is more efficient than in small environments. The variations in small-scale environments tend to make large augmentation ratios less effective for performance improvement, as can be observed from the slopes in both Fig. (\ref{fig:dataset_low}) and Fig. (\ref{fig:dataset_med}). Since the WILD2 Env. 2 dataset has only 3000 training samples, it is not included in Fig.(\ref{fig:dataset_med}).

    \subsection{Scenarios} \label{sec:scenarios}

        \subsubsection{Dataset Partition, Where to Sample?} \label{sec:partition}
            
\begin{figure}[h!]
    \centering
    \vspace{-5mm}
    \includegraphics[width =\columnwidth]{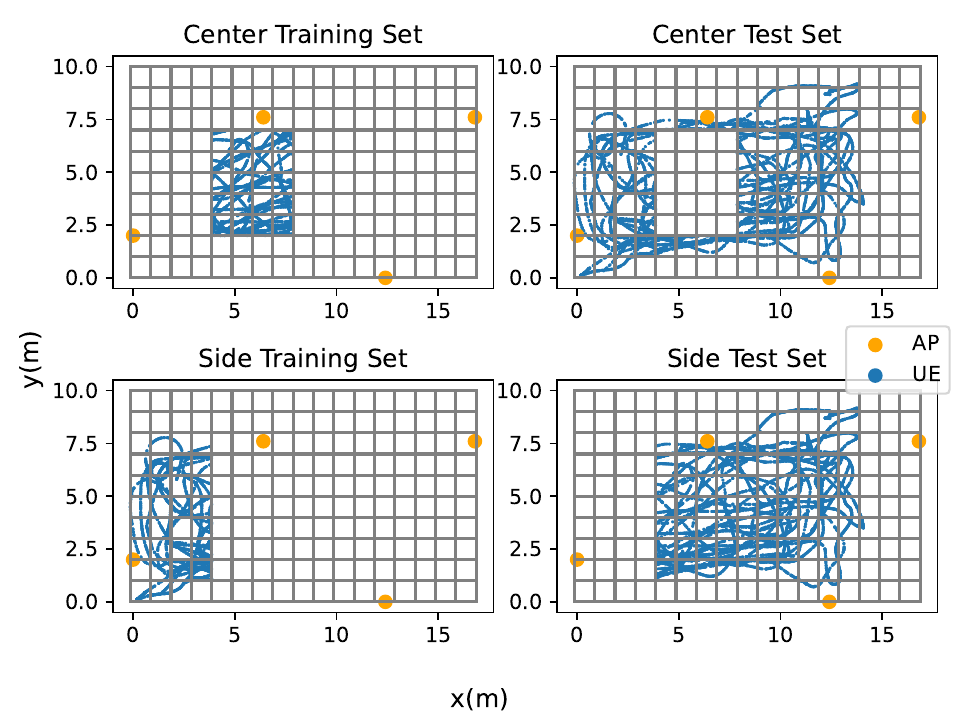}
    \caption{Maps of Training and Testing Datasets for Exclusive Data Partition}
    \label{fig:partition_map}
\end{figure}

The training datasets used in this work and the literature assume that the environment is sampled uniformly since uniform sampling of locations has significant advantages for ML-based localization. It is well known from classical sampling theory that (for a fixed number of samples) uniform sampling \emph{usually} provides the best reconstruction accuracy \cite{feichtinger}. Since learning localization 
mapping from features to locations becomes an interpolation problem as any test sample can be found between some samples in the training set, it stands to reason that uniform sampling is also best for ML-based localization. Moreover, the diversity of the data helps in learning the channel dynamics of the environment since indoor environments have scatterers distributed non-uniformly. However, in reality, it may not be possible to sample the environment in such a manner due to time and access restrictions.


In addition, we would like to generalize to extrapolation cases where there is no overlap between the areas containing the locations of the test and training set and demonstrate augmentation performance in such cases. Thus, we train over either a side of the room or the center or the side of the room and test it in the rest. Fig. (\ref{fig:partition_map}) shows the difference between the two schemes' training and sets.  

\begin{figure}[h!]
    \centering
    \vspace{-4.5mm}
    \includegraphics[width =\columnwidth]{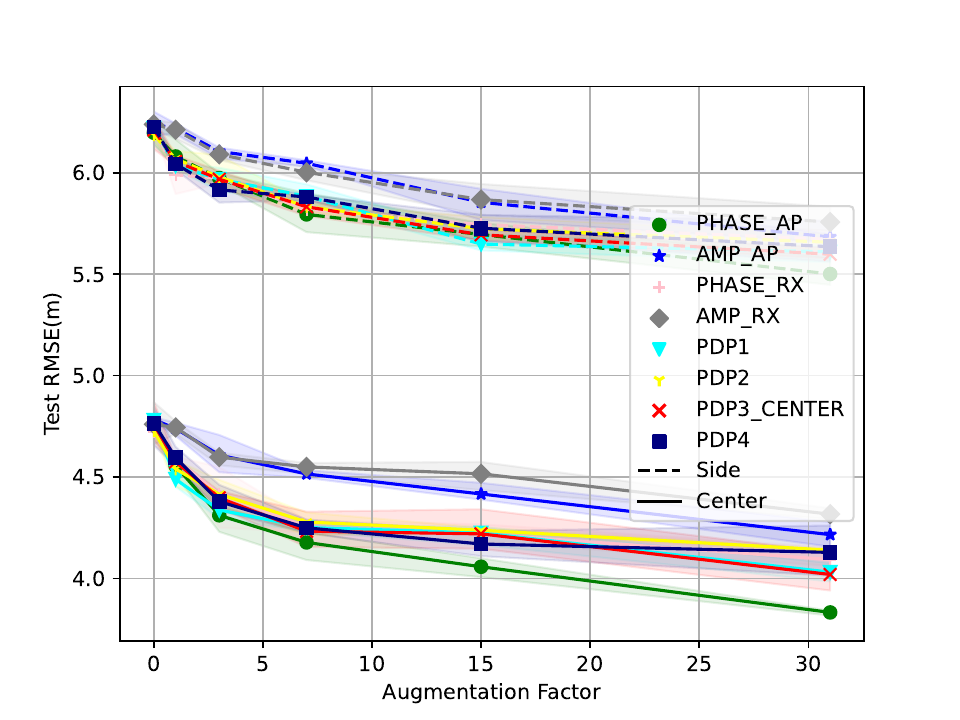}
    \caption{Test Set Performance vs Augmentation Size and Methods, Model: Fully Connected, Dataset: WILD1 Env. 2, Original Dataset Size: 4000}
    \label{fig:dataset_partition}
    \vspace{-3mm}
\end{figure}

Fig. (\ref{fig:dataset_partition}) shows the effect of selecting training points from such separated regions, and generally, training with samples in the center of the room results in significantly better performance than training on the sides - with the difference in RMSE of about $1.5$ m. Augmentation is also more effective for the sampling in the center, both in terms of absolute and relative improvement of the RMSE. Furthermore, we see that the different augmentation methods show a larger spread for the center training (up to 0.5 m) compared to the edge training (up to 0.3 m). Still, in both cases, methods that add phase fluctuations at the transceiver yield the best performance. 
Moreover, \ac{DL} methods suffer significantly in performing extrapolation compared to the interpolation cases (Sec. VI. C.). 

        \subsubsection{Sample Quality, Where to Augment?} \label{sec:quality}

\begin{figure}[h!]
    \centering
    \vspace{-5mm}
    \includegraphics[width =\columnwidth]{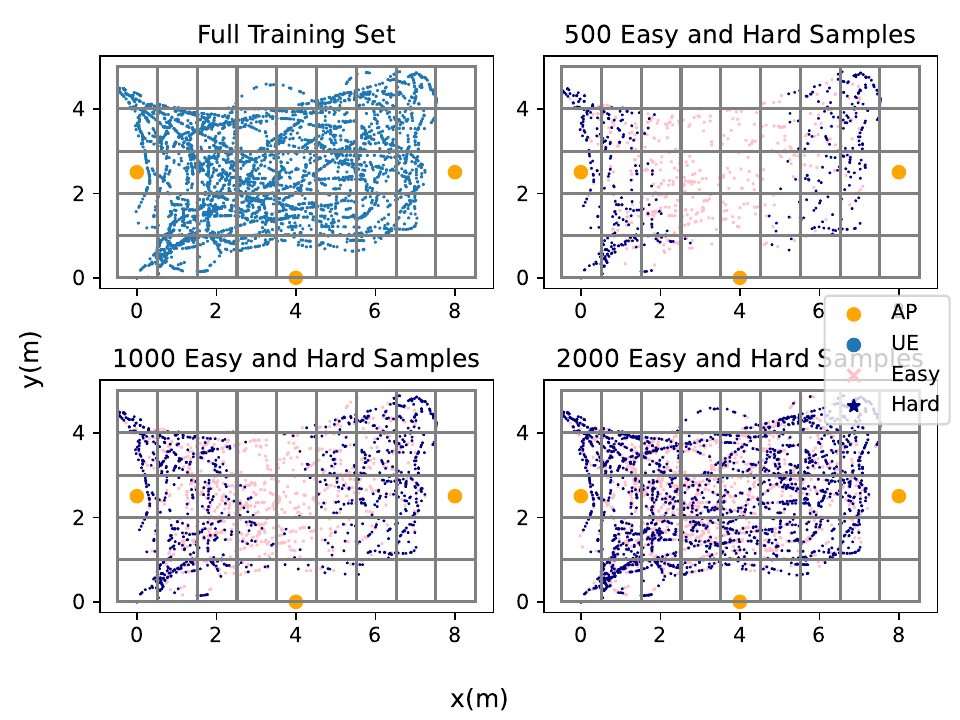}
    \caption{Distribution of Easy and Hard Samples, Dataset: WILD1 Env. 1}
    \label{fig:easy_hard_map}
\end{figure}

\begin{figure}[h!]
    \centering
    \vspace{-4.5mm}
    \includegraphics[width =\columnwidth]{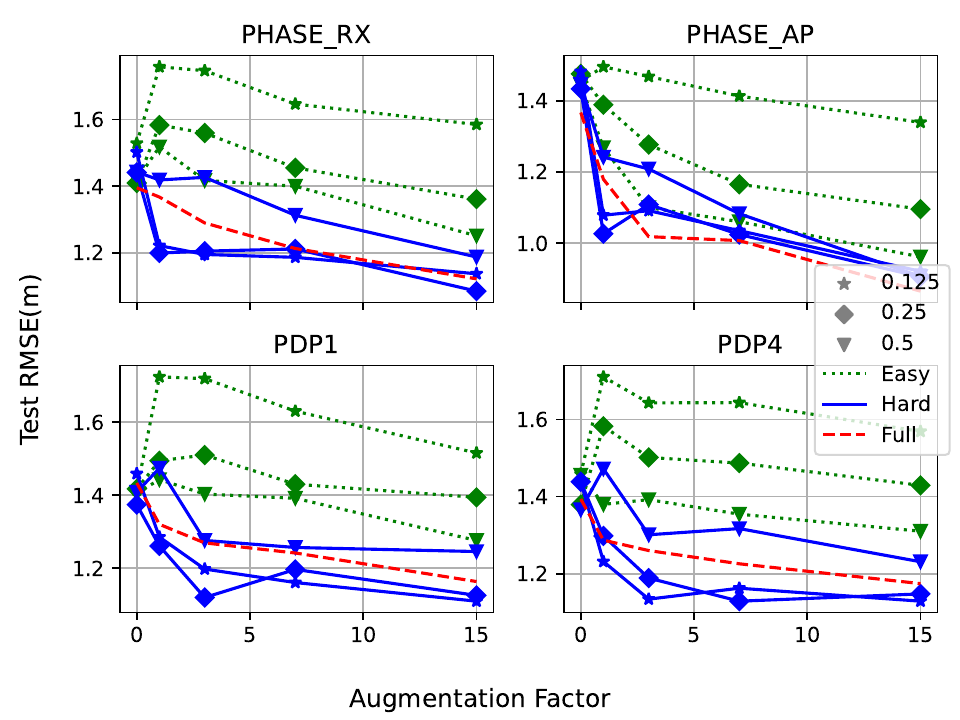}
    \caption{Test Set Performance vs Augmentation Size and Methods, Model: Fully Con., Dataset: WILD1 Env. 1, Original Dataset Size: 2000}
    \label{fig:paper_easyhard_fully}
\end{figure}

\begin{figure}[h!]
    \centering
    \vspace{-4mm}
    \includegraphics[width =\columnwidth]{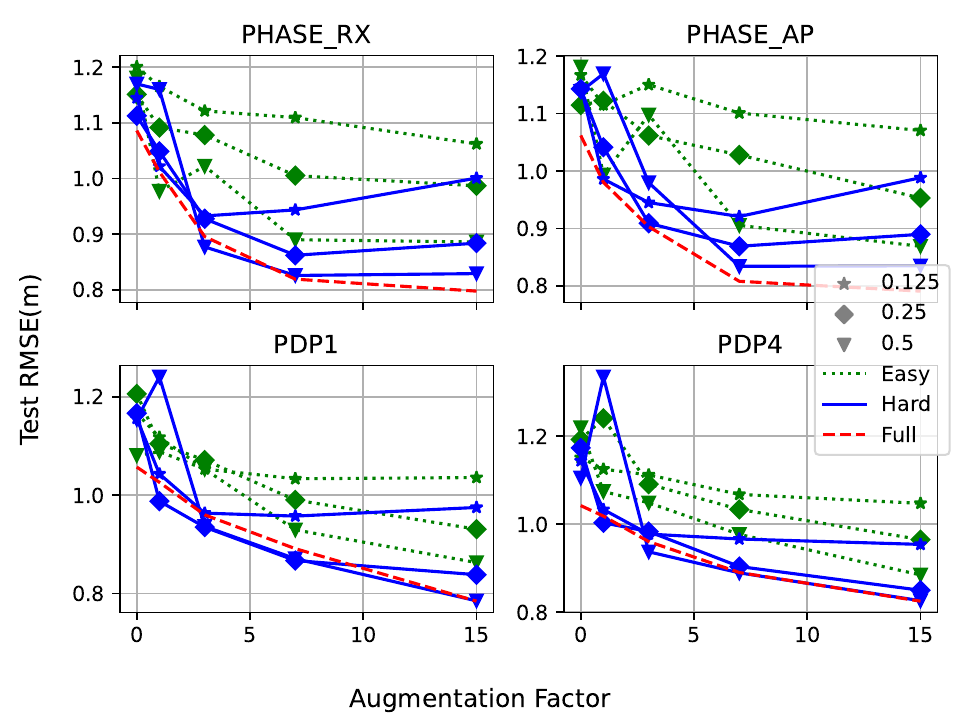}
    \caption{Test Set Performance vs Augmentation Size and Methods, Model: CNN, Dataset: WILD1 Env. 1, Original Dataset Size: 2000}
    \label{fig:paper_easyhard_cnn}
    \vspace{-3mm}
\end{figure}

The previous subsection studied where to sample for training points if we were given an environment for localization. Also, up to now we have augmented all the training samples equally. This subsection considers the case where we are already given a training dataset, and we aim to determine whether it is advantageous to augment some samples more than others, and if yes, which ones.  For this purpose, we calculate the average training loss for given sample $i$ as $\mathcal{L}_i \triangleq \frac{1}{E} \sum_{j = 1}^E \ell(f(\boldsymbol{x}_i; \theta_j), \boldsymbol{y}_i)$, where $E$ is the total number of epochs and $\theta_j$ is the model parameters after running the epoch $j$. Then, we classify samples as \textit{easy} if the average training loss is low; otherwise,  as \textit{hard}. The fraction of the samples we classify as ''hard" is a manually chosen parameter $\rho_{\rm hs}$. For example, if there are 100 training samples, they are ordered with respect to their average training loss, and we may use only the first $100 \times \rho_{\rm hs}$ as hard samples for the augmentation. In Fig. (\ref{fig:easy_hard_map}), we show the distinction between \textit{easy} and \textit{hard} samples in WILD 1 Env. 1, where \textit{easy} are more concentrated on the center of the room and \textit{hard} tend to be found on the sides. However, when we increase the number of selected samples, both classes become similar and cover the whole environment of interest. 
Fig.  (\ref{fig:paper_easyhard_fully}) and Fig. (\ref{fig:paper_easyhard_cnn}) present results for three different ratios of hard samples, i.e., $\rho_{\rm hs}=0.5,$ 0.25, 0.125, both for CNN and FCNN. We augment only the \textit{hard} samples, and for fairness of the comparison, ensure that the number of samples generated per data point is $1/\rho_{\rm hs}$ so that the total number of augmented samples is independent of $\rho_{\rm hs}$. 
Results show that augmenting the \textit{hard} samples might improve performance compared to augmenting \textit{easy} samples. The difference in performance via \textit{easy} and \textit{hard} samples is more pronounced when the number of selected samples is smaller, i.e., small $\rho_{\rm hs}$, for FCNN in Fig. (\ref{fig:paper_easyhard_fully}). Moreover, augmenting only \textit{hard} samples yields better performance than the case of augmenting the entire dataset. 
In Fig. (\ref{fig:paper_easyhard_cnn}), which uses a CNN, augmenting \textit{hard} samples provides better performance. However, in contrast to FCNN, increasing the selection ratio improves localization accuracy. Similar to the FCNN case, the performance of the full augmentation can be reached by augmenting only a portion of the data, namely the \textit{hard} samples. This is intuitive, too,  since easy samples are associated with near-zero-loss; augmented data created from them also have near-zero-loss, so that their inclusion in the training process results in negligible improvement of the localization accuracy. 
Thus, by running the dataset for a training algorithm once and determining the training loss for the different samples, we can learn where to augment it in a given environment and dataset. 

        \subsubsection{Transfer Learning} \label{sec:transfer_learning}
            \begin{figure}[h!]
    \centering
    \vspace{-4.5mm}
    \includegraphics[width =\columnwidth]{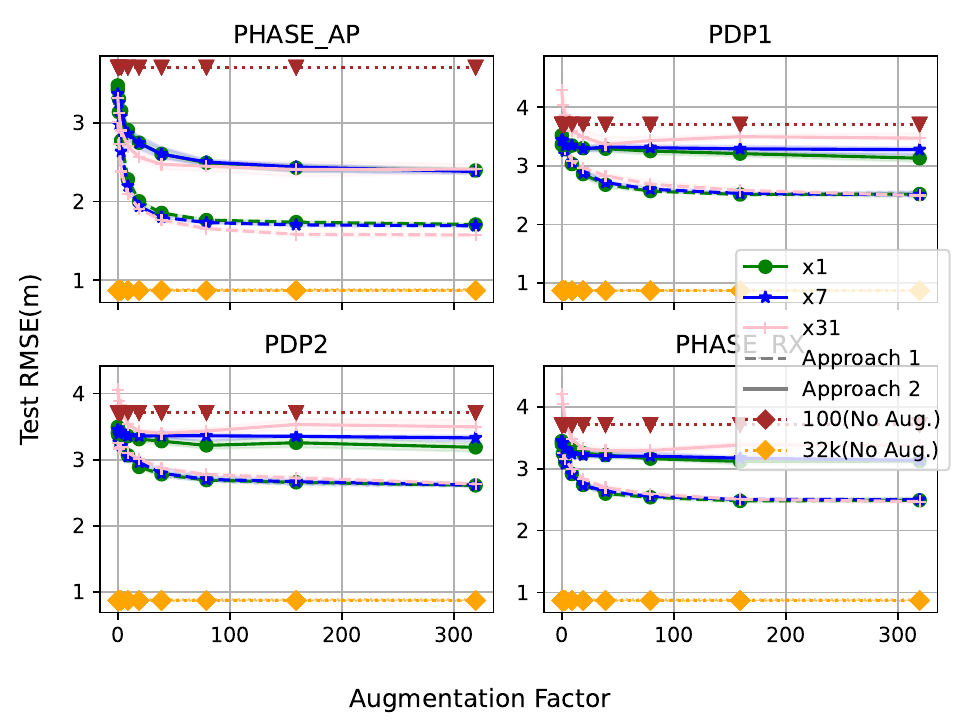}
    \caption{Test Set Performance vs Augmentation Size and Methods, Datasets: WILD1 Env. 1 to WILD1 Env. 2, Model: Fully C., Original Source Dataset Size: 1000, Original Target Dataset Size: 100}
    \label{fig:paper_tl_low_wild1}
\end{figure}

 \ac{TL} can significantly reduce the data collection burden.  \ac{TL} is training a model that is data rich in the source domain $(\mathcal{X}_{\mathrm{S}}, \mathcal{Y}_{\mathrm{S}})$ (in the following example WILD 1 Env. 1) and using it in a low data environment called target domain $(\mathcal{X}_{\mathrm{T}}, \mathcal{Y}_{\mathrm{T}})$, WILD 1 Env. 2. 
 We assume that both feature and label spaces are the same between WILD 1 Env. 1 and WILD 1 Env. 2. However, underlying mappings $f_{\mathrm{S}}$ and $f_{\mathrm{T}}$ are different due to them being two different environments. We model the \ac{NN} as follows: $f(\bold{x}) = g(\phi(\bold{x}))$, where $g(\cdot)$ is the task-specific last layers and $\phi(\cdot)$ is the first layer of the \ac{NN}, i.e., feature extractor. We follow two approaches, \textit{i)} Updating {\em all} the parameters in the fine-tuning process with the same optimization factors and using target domain data only, \textit{ii)} Freezing the feature extractors while re-training the last layers from scratch in the target domain. In CNN, the feature extractor part is the convolutional layers (see Sec. VI. A). In FCNN, the first two fully connected layers are used as the feature extractor. Similarly, task-specific layers are the last layers of the corresponding networks.

\begin{figure}[h!]
    \centering
    \vspace{-5mm}
    \includegraphics[width =\columnwidth]{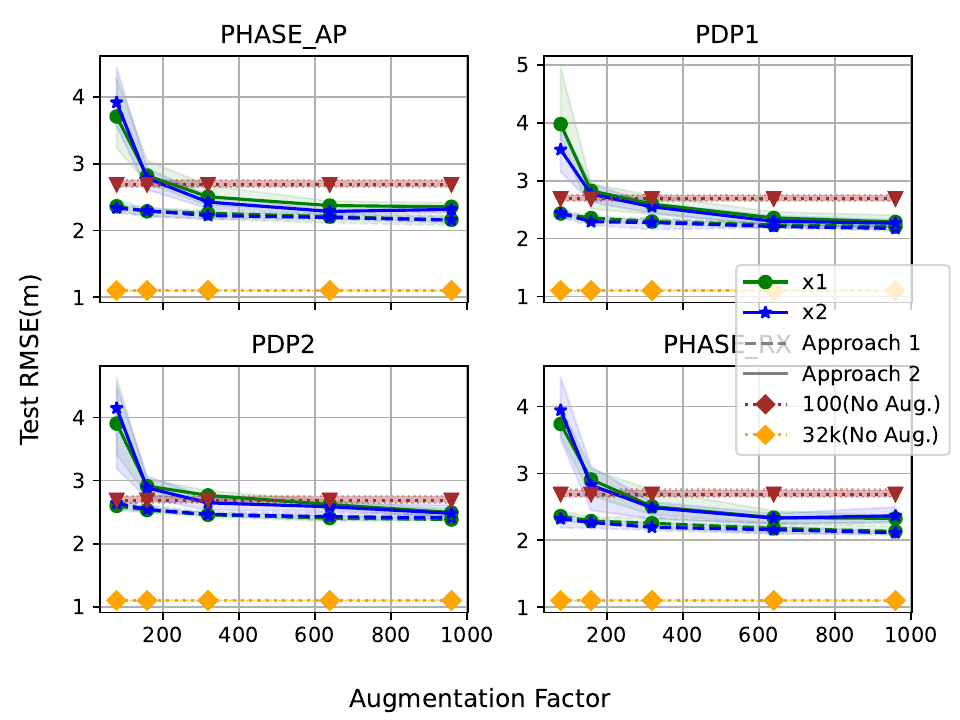}
    \caption{Test Set Performance vs Augmentation Size and Methods, Datasets: WILD1 Env. 1 to WILD1 Env. 2, Model: 
    CNN, Original Source Dataset Size: 32000, Original Target Dataset Size: 100}
    \label{fig:paper_tl_high_wild1}
    \vspace{-3mm}
\end{figure}

Fig. (\ref{fig:paper_tl_low_wild1}) presents the results of the two TL approaches for a case where the source domain has 1000 and the target domain has 100 samples. Fig. (\ref{fig:paper_tl_low_wild1}) evaluates augmentation factors in the source domain with 3 different multipliers ($\times 1, \times 7, \times 31$), and the target domain is augmented to up to 32000 samples. We consider a similar scenario in Fig. (\ref{fig:paper_tl_high_wild1}) where the source domain has 32000 measurements in the dataset, with 100 measurements in the target domain. We also provide the results for dataset sizes of 100 and 32000 without any augmentation. 

The results provide four main takeaways: \textit{i)} using a very large source dataset is not beneficial, as it may learn the source model well, which makes the performance in the target domain worse; \textit{ii)} updating all of the \ac{NN} parameters in the target domain provides better performance than freezing the feature extractor and re-training the  last layers is more useful, \textit{iii)}  target domain augmentation is very beneficial, providing 1 meter, i.e., $30$ \% of RMSE reduction in our examples, \textit{iv)} source domain augmentation is not beneficial if target domain augmentation is performed.

\section{Conclusion} \label{sec:conclusion}
    
In this work, we proposed DA methods that utilize domain knowledge to mitigate the problem of laborious data collection in \ac{DL}-based indoor localization. We observed that channel-based augmentation methods perform better when the data regime is low. Transceiver-based methods work well in medium and high data regimes as the channel itself is learned properly by the model with a large amount of data. It is highlighted that the overall performance is better when the dataset contains more real measurements. Large and NLOS environments benefit from DA more than small and LOS environments. Finally, we provided a full augmentation {\em strategy} starting from where to sample a given environment, which samples to augment, and how to augment the dataset during the \ac{DA}. We concluded that sampling the center portion of the environment generally provides better results than at the edges of the environment and that augmenting hard samples provides better localization accuracy. Finally, in \ac{DA}, target domain augmentation is crucial, while source domain augmentation does not significantly affect performance in target domains. Overall, these strategies and augmentation methods allow us to significantly reduce the data collection effort for ML-based localization methods without sacrificing accuracy. 

\bibliographystyle{IEEEtran}
\bibliography{bibli}
\end{document}